\documentclass[onecolumn,showpacs]{revtex4}

\usepackage{graphicx}
\usepackage{dcolumn}
\usepackage{bm}

\input epsf

\begin{document}

\title{\Large Tolman-Bondi Space-Time in Brane World Scenario}

\author{\bf Subenoy
Chakraborty$^1$\footnote{subenoyc@yahoo.co.in}, ~Asit
Banerjee$^2$\footnote{asitb@cal3.vsnl.net.in} and Ujjal
Debnath$^3$\footnote{ujjaldebnath@yahoo.com}}

\affiliation{$^1$Department of Mathematics, Jadavpur University,
Calcutta-32, India.\\ $^2$Department of Physics, Jadavpur
University, Calcutta-32, India.\\ $^3$Department of Mathematics,
Bengal Engineering and Science University, Shibpur, Howrah-711
103, India.\\}

\date{\today}

\begin{abstract}
In the present work, inhomogeneous Tolman-Bondi type dust
space-time is studied on the brane. There are two sets of
solutions of the above model. The first solution represents either
a collapsing model starting from an infinite volume at infinite
past to the singularity or a model starting from a singularity and
expanding for ever having a transition from decelerating phase to
accelerating phase. The first solution shows that the end state
of collapse may be black hole or a naked singularity depending
signs of various parameters involved. The second solution
represents a bouncing model where the bounce occurs at different
comoving radii at different epochs.

\end{abstract}

\pacs{04.20~-q,~~04.40~ Dg,~~97.10.~CV.}

\maketitle

In the seminal papers [1, 2] the introduction of branes into
cosmology offered a novel approach to the understanding  of the
evolution of our universe. It was proposed that our 4D universe is
said to be a singular hypersurface, a 3-brane embedded in a
$(4+d)$ dimensional space-time, which is called the bulk. Here in
the bulk, we have considered no matter field. The matter field is
all confined to the brane and the gravity propagates in the extra
dimensions as well. The effective equations for gravity in four
dimensions were obtained by Shiromizu et al [3] and discussed by
Roy Maartens [4]. Here Israel's boundary conditions were used
along with $z_{2}$ symmetry for the five dimensional bulk
space-time embedding our brane universe at a fixed point
$y=y_{0}$. If we add a cosmological constant in the bulk, the
solutions of the Einstein's equations
$G_{ab}=\kappa_{5}^{2}T_{ab}$ can be obtained, where the universe
starts with a non-conventional phase and then enters the standard
cosmological phase [5 - 11]. Friedmann like equations are usually
solved on the brane for matter localized on the brane [12]. The
usual case in the brane model recollapses if it is embedded in an
anti de-Sitter ($\Lambda_{5}<0$) five dimensional bulk, where the
extra dimension is space-like. In a recent work of Ponce de Leon
[13] the acceleration of the present universe is studied as a
consequence of the time evolution of the vacuum energy in
cosmological models based on brane world theories in 5
dimensions. The basic two assumptions in this paper are that the
universe is spatially flat and the matter density decrease as an
inverse power of the scale factor. If $\Lambda_{5}=0$ then the
induced cosmological term in 4D is positive and the brane
universe expands continuously in a power law time dependence.
Again for $\Lambda_{5}>0$, this universe becomes dominated by a
positive cosmological term $\Lambda_{4}$. The effect is an
asymptotic de-Sitter expansion which occurs regardless of the
signature of the extra dimension. On the otherhand, with a single
timelike extra dimension the cosmological brane world model
experiences a natural bounce without ever-reaching the singular state [14].\\

In the present work, we have considered the Tolman-Bondi
space-time on the brane. This is an inhomogeneous and anisotropic
space-time and differs from the homogeneous Friedmann space-time
so far mentioned above regarding their dynamical behaviour. It is
perhaps for the first time that Tolman-Bondi type dust space-time
is studied on the brane. We'll consider only the spatially flat
model ($f(r)=0$), which is consistent with the present
observational data. The projection of the bulk Weyl tensor on the
brane is also assumed to be zero. There occur two different cases
so long as we assume the anti de-Sitter bulk and the extra
dimension to be space-like. In the case I, the brane world
collapses from an infinite past to a singularity at a finite
epoch or otherwise begins from the singularity and expands
indefinitely with deceleration near the big bang and acceleration
at the late stage. In case II, the brane world model has a lower
bound and hence singularity-free, similar to the case mentioned
above (Ref. [14]). Here we get a contracting model bouncing from
a lower bound, which however takes place at different shells with
different comoving radii at different epochs. This is usual for an
inhomogeneous space-time. It is to be noted further that in the
Tolman-Bondi space-time on brane under consideration, there is no
recollapsing model.\\

{\bf Field equations in the Brane:}\\

The modified 4D Einstein's equations on the brane embedded in the
5D bulk are usually presented in the form [1, 3]

\begin{equation}
G_{\mu\nu}=-\Lambda_{4}g_{\mu\nu}+\kappa_{4}^{2}T_{\mu\nu}+\kappa_{5}^{4}S_{\mu\nu}-E_{\mu\nu}
\end{equation}

where $\Lambda_{4}$ is the effective four dimensional
cosmological constant, $\kappa_{5}$ and $\kappa_{4}$ are the
coupling constants related to the gravitational constants in 5D
bulk and 4D brane respectively. They are further related with the
fundamental 5D Planck mass and the effective Planck mass on the
brane. $T_{\mu\nu}$ is the usual matter energy tensor on the
brane. The two additional terms $S_{\mu\nu}$ and $E_{\mu\nu}$ on
the r.h.s of (1) stand  respectively for the local quadratic
energy-momentum correction term and the free energy induced on
the brane by the 5D Weyl tensor in the bulk. The local correction
term arises from the extrinsic curvature of the brane. Their
explicit expressions are given by

\begin{equation}
S_{\mu \nu}=\frac{1}{12}T~ T_{\mu\nu}-\frac{1}{4}T^{\alpha}
_{\mu}T_{\nu \alpha}+\frac{1}{24} g_{\mu \nu}(3 T^{\alpha \beta }
T_{\alpha \beta }-T^{2})
\end{equation}

Finally, $E_{\mu\nu}$ in (1) is the projection of the bulk Weyl
tensor on the brane defined by

\begin{equation}
E_{ab}=C_{a b c d} n^{c} n^{d}
\end{equation}

Here $C_{a b c d}$ is the usual Weyl tensor in the bulk, $n^{a}$
is the unit normal to the hypersurface $y=$ constant,
representing the brane world, where $y$ represents the extra
fifth dimension. In the above the latin indices refer to the 5D
bulk and greek indices correspond to the brane world. The four
dimensional $\Lambda_{4}$ and 5D cosmological constant
$\Lambda_{5}$ [2, 3] are related by the following relation

\begin{equation}
\Lambda_{4} =\frac{\kappa^{2}_{5}}{2} (\Lambda _{5}+ \frac{\kappa
^{2} _{5} \lambda ^{2} }{6})
\end{equation}

where $\lambda$ is the so-called brane tension or the vacuum
energy of the brane world. This brane tension is further related
to the 4D gravitational constant by a relation of the form

\begin{equation}
\kappa_{4}^{2} =\frac{\kappa ^{4} _{5}  \lambda}{6}
\end{equation}

In this work, we assume that our brane model embedded at $y=$
constant in the conformally flat bulk (zero Weyl tensor) is in
homogeneous and an isotropic. As a result the term $E_{\mu\nu}$ is
neglected and is justified in view of the late time dominance of
the vacuum energy compared to the brane matter density $\rho$
i.e., $\lambda >> \rho$ [4] being compatible with observations.
Using spherically symmetric form the metric is given by

\begin{equation}
ds^{2}=dt^{2}-e^{\alpha}dr^{2}-R^{2}d\Omega^{2}
\end{equation}

where $\alpha=\alpha(r,t)$ and $R=R(r,t)$.\\

Now considering matter in the form of dust on brane the
independent field equations are given by

\begin{equation}
\frac{1}{R^{2}}-e^{-\alpha}\left(2~\frac{R''}{R}+\frac{R'^{2}}{R^{2}}-\alpha'~\frac{R'}{R}
\right)+\frac{\dot{R}^{2}}{R^{2}}+\dot{\alpha}~\frac{\dot{R}}{R}=\Lambda+\rho+\frac{\rho^{2}}{2\lambda}
\end{equation}

\begin{equation}
\frac{1}{R^{2}}-e^{-\alpha}~\frac{R'^{2}}{R^{2}}+2~\frac{\ddot{R}}{R}+\frac{\dot{R}^{2}}{R^{2}}
=\Lambda-\frac{\rho^{2}}{2\lambda}
\end{equation}

\begin{equation}
-e^{-\alpha}\left(\frac{R''}{R}-\frac{\alpha'}{2}\frac{R'}{R}
\right)+\frac{\ddot{\alpha}}{2}+\frac{\dot{\alpha}^{2}}{4}+\frac{\dot{\alpha}}{2}\frac{\dot{R}}{R}+\frac{\ddot{R}}{R}
=\Lambda-\frac{\rho^{2}}{2\lambda}
\end{equation}
and
\begin{equation}
2~\frac{\dot{R}'}{R}-\dot{\alpha}~\frac{R'}{R}=0
\end{equation}

It follows immediately from the equation (10)

\begin{equation}
e^{\alpha}=\frac{R'^{2}}{1+f(r)}
\end{equation}

In general there may be three different choices
$f(r)=0,~f(r)>0,~f(r)<0$. For simplicity $f(r)$ is assumed to be
zero (marginally bound case). In this particular case the field
equations (7) to (9) simplify to

\begin{equation}
\frac{\dot{R}^{2}}{R^{2}}+2~\frac{\dot{R}}{R}\frac{\dot{R}'}{R'}=\Lambda+\rho+\frac{\rho^{2}}{2\lambda}
\end{equation}

\begin{equation}
2~\frac{\ddot{R}}{R}+\frac{\dot{R}^{2}}{R^{2}}=\Lambda-\frac{\rho^{2}}{2\lambda}
\end{equation}

\begin{equation}
\frac{\ddot{R}'}{R'}+\frac{\dot{R}}{R}\frac{\dot{R}'}{R'}+\frac{\ddot{R}}{R}=\Lambda-\frac{\rho^{2}}{2\lambda}
\end{equation}

It is to be noted that the energy density enters the field
equations quadratically in contrast with the usual general
relativistic 4D equations.\\

Subtracting (13) from (14) we get

\begin{equation}
\frac{\ddot{R}'}{R'}+\frac{\dot{R}}{R}\frac{\dot{R}'}{R'}-\frac{\ddot{R}}{R}-\frac{\dot{R}^{2}}{R^{2}}=0,
\end{equation}

which can be integrated twice to yield

\begin{equation}
\dot{R}^{2}=\mu R^{2}+\frac{D(r)}{R}
\end{equation}

In (16), $\mu$ is in general a function of time and $D(r)$ is a
function of radial co-ordinate alone. In order to integrate
equation (16) we assume further $\mu=$ constant, which in fact has
two groups of solutions. Also from the field equations (12) - (14)
using (16), we finally get

\begin{equation}
\rho=6\mu-2\Lambda+\frac{D^{2}(r)}{R^{2}R'}
\end{equation}
and
\begin{equation}
\frac{\rho^{2}}{2\lambda}=\Lambda-3\mu.
\end{equation}

{\bf Solutions:}\\

There are three possible types of solutions of the evolution
equation (16) depending on the signs of the parameter $\mu$ and
the function $D(r)$ namely,
$(i)~~\mu>0,~D(r)>0,~~(ii)~~\mu>0,~D(r)<0,~~(iii)~~\mu<0,~D(r)>0$.
However, the third possibility will not be discussed as it gives
oscillatory solution, not of much interest in the present
context.\\

{\bf Case~I~:} $\mu>0,~D(r)>0 :$\\

The solution for the scale factor $R$ from equation (16) gives

\begin{equation}
R=R_{0}(r)~Sinh^{2/3}u
\end{equation}

where,
$R_{0}^{3}=\frac{D(r)}{\mu},~u=a(t_{0}(r)-t),~a=\sqrt{\mu}$ and
$t_{0}(r)$ is an arbitrary function of the radial co-ordinate
$r$.\\

Differentiating $R$ we get in turn

\begin{equation}
\dot{R}=-\frac{2aR_{0}Coshu}{3Sinh^{1/3}u}~~\text{and}~~\ddot{R}=\frac{2a^{2}R_{0}[Cosh2u-2]}{9Sinh^{4/3}u}
\end{equation}

Here there may be two situations. The first case corresponds to
$t<t_{0}$. The collapse starts from the infinite past
($t\rightarrow -\infty$) i.e., when $R\rightarrow \infty$ and
finally converges to the singularity $R\rightarrow 0$ at the time
$t_{0}(r)$. The rate of contraction at both ends is infinitely
large. It is a collapsing model of the brane. One should note
however that different shells collapse at different epochs
$t_{0}(r)$ because of inhomogeneity. The time of formation of
central singularity is given by [15 - 17]

\begin{equation}
t_{s}=t_{i}+\frac{1}{a}~
\begin{array}{c}
lim\\
{r\rightarrow 0}
\end{array}
Sinh^{-1}\left(\frac{\mu r^{3}}{D(r)} \right)^{1/2}
\end{equation}

where we have chosen $R=r$ at $t=t_{i}$ as an initial condition.\\

Now at the time of formation of apparent horizon $(t_{ah}(r))$ for
a shell having comoving radial co-ordinate $r$, we must have
$\dot{R}^{2}=1$. Using dynamical equation (16) it results

\begin{equation}
\mu R^{3}-R+D(r)=0.
\end{equation}

Thus the time difference between the formation of apparent
horizon and the central singularity is [15 - 17]

\begin{equation}
t_{ah}(r)-t_{s}=\frac{1}{a}~\left[ Sinh^{-1}\left(\frac{r}{R_{0}}
\right)^{3/2}-Sinh^{-1}\left(\frac{R_{ah}}{R_{0}} \right)^{3/2}
\right]-\frac{1}{a}~
\begin{array}{c}
lim\\
{r\rightarrow 0}
\end{array}
Sinh^{-1}\left(\frac{\mu r^{3}}{D(r)} \right)^{1/2}
\end{equation}

where $R_{ah}$ is the positive real root of the above cubic
equation (22) in $R$. Further, in order to have the collapsing
process starts from a regular initial data one should have the
density to be finite at $r=0$ when $t=t_{i}$~. Thus near $r=0$ the
function $D(r)$ should have the power series expansion

\begin{equation}
D(r)=D_{0}r^{3}+D_{1}r^{4}+D_{2}r^{5}+... ...
\end{equation}

\begin{figure}
\includegraphics[height=2.7in]{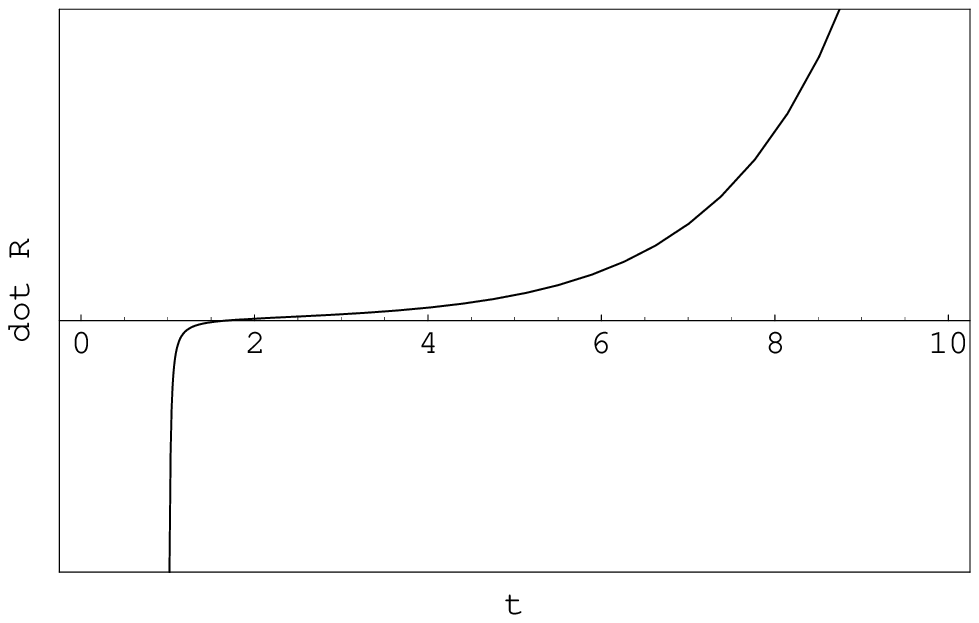}\\
\vspace{1mm} Fig.1\\

\vspace{5mm} Fig. 1 shows the behaviour of $\ddot{R}$ of eq.(20)
with time in case I for $\mu=1, ~D_{0}=1$ and $t_{0}=1$.
\hspace{2cm} \vspace{4mm}
\end{figure}

\begin{figure}
\includegraphics[height=2.7in]{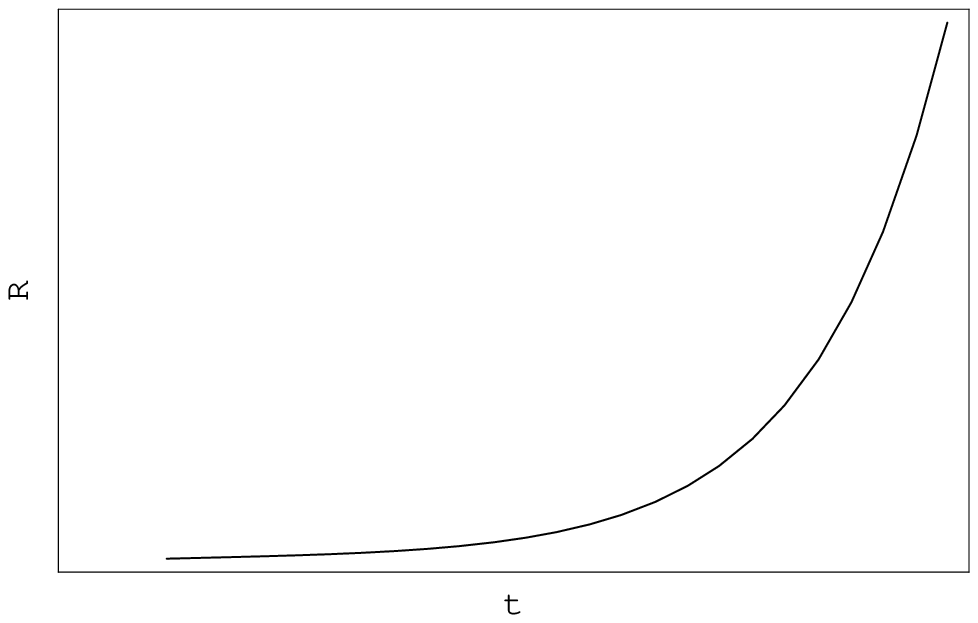}\\
\vspace{1mm} Fig.2\\

\vspace{5mm} Fig. 2 shows the behaviour of $R$ of eq.(26) with
time in case II for $\mu=1, ~R_{0}=1$ and $t_{0}=1$. \hspace{2cm}
\vspace{4mm}
\end{figure}

Then the above time difference simplifies to

\begin{equation}
t_{ah}(r)-t_{s}=-\frac{D_{1}}{2D_{0}\sqrt{\mu+D_{0}}}~r+\frac{(2\mu+3D_{0})D_{1}^{2}-4D_{0}D_{2}(\mu+D_{0})
}{8D_{0}^{2}(\mu+D_{0})^{3/2} }~r^{2}+O(r^{3})
\end{equation}

Thus we note that the sign of this time difference depends on
various parameters involved. Therefore, the spherically symmetric
space-time shows collapse to a black hole or a naked singularity
near the centre depending on the nature and properties of certain
parameters.\\

On the otherhand, if $t\ge t_{0}$, $R(t)$ starts from the
singularity $R=0$ at $t=t_{0}$ and expands indefinitely. Although
this shows a monotonic expansion $\ddot{R}<0$ initially and
subsequently changes sign (see figure 1) to $\ddot{R}>0$ at the
later stage. Thus there is a transition from deceleration to
acceleration during the evolution of the brane model. This result
seems to be interesting because the model is consistent with the
possibility of the structure formation in the early stage
$(\ddot{R}<0)$ and the accelerated expansion in the later stage
getting support from the observational data.\\

{\bf Case II :} $\mu>0,~D(r)<0$:\\

In view of (16) this case is restricted by the condition that
$\mu R^{3}>|D(r)|$. The solutions are given by

\begin{equation}
R=R_{0}(r)Cosh^{2/3}u
\end{equation}

Hence,
\begin{equation}
\dot{R}=-\frac{2aR_{0}Sinhu}{3Cosh^{1/3}u}~~\text{and}~~\ddot{R}=\frac{2a^{2}R_{0}[Cosh2u+2]}{9Cosh^{4/3}u}
\end{equation}

In this case, $R$ is never zero, which shows that the brane world
model is singularity free. This conclusion follows immediately
from the positive value of $\ddot{R}$ throughout the evolution
and hence there is a minimum only and no maximum at any stage.
$R$ has an infinite magnitude at $t\rightarrow -\infty$,
subsequently contracts to $R=R_{0}$ when $t=t_{0}$ and then again
explodes to infinity at $t\rightarrow \infty$ (see figure 2). So
this particular brane world model shows bounce. However, it
should be noted that due to inhomogeneity in the energy
distribution the bounce occurs at different comoving radii at
different epochs.\\

{\bf References:}\\
\\
$[1]$ L. Randall and R. sundrum, {\it Phys. Rev. Lett.} {\bf 83} 4690 (1999)\\
$[2]$ L. Randall and R. sundrum, {\it Phys. Rev. Lett.} {\bf 83} 4690 (1999).\\
$[3]$ T. Shiromizu, K. Maeda and M. sasaki, {\it Phys. Rev. D} {\bf 62} 024012 (2000).\\
$[4]$ R. Maartens, {\it Geometry and Dynamics of the Brane World}
in: Pascual-Sanchez, J. (ed), Reference frames and
Gravito-magnetism, pp. 93-119, World Sci. 2001; gr-qc/0101059.\\
$[5]$ P. Binetruy, C. Deffayet, V. Ellwanger and D. Langlois,
{\it Phys. Lett. B} {\bf 477} 285 (2000).\\
$[6]$ N. Kaloper and A. Linde, {\it Phys. Rev. D} {\bf 59} 101303
(1999).\\
$[7]$ R. N. Mahapatra, A. Perez-Lorenzana and C. A. de Spires,
{\it Int. J. Mod. Phys. A} {\bf 16} 1431 (2001).\\
$[8]$ N. Kaloper, {\it Phys. Rev. D} {\bf 60} 123506 (1999).\\
$[9]$ P. Kanti, I. I. Kagon, K. A. Olive and M. Pospelov, {\it
Phys. Lett. B} {468} 31 (1999).\\
$[10]$ G. Dvali and S. H. H. Tye, {\it
Phys. Lett. B} {\bf 450} 72 (1999).\\
$[11]$ E. E. Flanagan, S. H. H. Tye and I. Wasserman, {\it
hep-th}/9909373.\\
$[12]$ J. Ponce de Leon, {\it Class. Quantum Grav.} {\bf 20} 5321
(2003).\\
$[13]$ J. Ponce de Leon, {\it Gen. Rel. Grav.} {\bf 37} 53
(2005).\\
$[14]$ Y. Shtanov and V. Sahni, {\it Phys. Lett. B}, IUCAA
pre-print-19/2003.\\
$[15]$  A. Banerjee, U. Debnath and S. Chakraborty, {\it Int. J. Mod. Phys. D} {\bf 12} 1255 (2003).\\
$[16]$ P. S. Joshi, N. Dadhich and R. Maartens, {\it Phys. Rev.
D} {\bf 65} 101501(R) (2002).\\
$[17]$ P.S. Joshi, \textit{Global Aspects in Gravitation and
Cosmology, }(Oxford Univ. Press, Oxford, 1993).\\

\end{document}